\documentclass[preprint, sort&compress]{elsarticle}

\usepackage{graphicx}% Include figure files
\expandafter\ifx\csname package@font\endcsname\relax\else
 \expandafter\expandafter
 \expandafter\usepackage
 \expandafter\expandafter
 \expandafter{\csname package@font\endcsname}%
\fi

\begin{document}

\title{Probing maximum energy of cosmic rays in SNR through gamma rays and neutrinos from the molecular clouds around SNR W28}

\author[nbu]{Prabir Banik}
\ead{pbanik74@yahoo.com}

\author[nbu]{Arunava Bhadra}
\ead{aru\_bhadra@yahoo.com}

\cortext[cor]{Arunava Bhadra}

%\author {Tamal Sarkar \thanks{Email address: ta.sa.nbu@hotmail.com}, Shubhrangshu Ghosh \thanks{Email address: shubhrang.ghosh@gmail.com}, Arunava Bhadra \thanks{Email address: aru\_bhadra@yahoo.com}} 

\address[nbu]{High Energy $\&$ Cosmic Ray Research Centre, University of North Bengal, Siliguri, WB, 734013, India}

%\author[Banik \& Bhadra]{Prabir Banik\thanks{Email address: pbanik74@yahoo.com} and Arunava Bhadra\thanks{Email address: aru\_bhadra@yahoo.com} \\
%High Energy $\&$ Cosmic Ray Research Centre, University of North Bengal, Siliguri, West Bengal, India 734013 }

%\affiliation{ High Energy $\&$ Cosmic Ray Research Centre, University of North Bengal, Siliguri, West Bengal, India 734013\\}
%\maketitle

%\label{firstpage}

\begin{abstract}
The galactic cosmic rays are generally believed to be originated in supernova remnants (SNRs), produced in diffusive shock acceleration (DSA) process in supernova blast waves driven by expanding SNRs. One of the key unsettled issue in SNR origin of cosmic ray model is the maximum attainable energy by a cosmic ray particle in the supernova shock. Recently it has been suggested that an amplification of effective magnetic field strength at the shock may take place in young SNRs due to growth of magnetic waves induced by accelerated cosmic rays and as a result the maximum energy achieved by cosmic rays in SNR may reach the knee energy instead of $\sim 200$ TeV as predicted earlier under normal magnetic field situation. In the present work we investigate the implication of such maximum energy scenarios on TeV gamma rays and neutrino fluxes from the molecular clouds interacting with the SNR W28. The authors compute the gamma-ray and neutrino flux assuming two different values for the maximum energy reached by cosmic rays in the SNR, from CR interaction in nearby molecular clouds. Both protons and nuclei are considered as accelerated particles and as target material. Our findings suggest that the issue of the maximum energy of cosmic rays in SNRs can be observationally settled by the upcoming gamma-ray experiment the Large High Altitude Air Shower Observatory (LHAASO). The estimated neutrino fluxes from the molecular clouds are , however, out of reach of the present/near future generation of neutrino telescopes.   
\end{abstract}

%\pacs{ 96.50.S-, 98.70.Rz, 98.70.Sa}
%\keywords{Cosmic rays, solar radiation, neutrinos, gamma rays}

%\maketitle
\maketitle

\section{Introduction}

There are some convincing arguments that cosmic rays with energy at least up to the ``knee" of the cosmic ray energy spectrum ($\sim 3$ PeV) are originated within our galaxy \cite{ber90}. Among the galactic sources, supernova remnants (SNRs) are believed to be the main source of cosmic rays (e.g. \cite{ama14, bla13}). The energy released in supernova explosions satisfies the energy requirement to maintain cosmic ray energy density in the galaxy considering an overall $\sim 10\%$ efficiency of the conversion of explosion energy into cosmic ray particles. The diffusive shock acceleration (DSA) at supernova blast waves driven by expanding SNRs can provide the necessary power law spectral shape of cosmic rays \cite{fer54, axf77, bla78, bel78, kry77, bel01}. 

The High Energy Stereoscopic System (HESS) collaboration recently reported the discovery of a $\gamma$-ray diffuse emission from a small region surrounding the galactic center (i.e. surroundings Sagittarius A* or Sgr A*) that extends up to $\sim 50$ TeV with no statistically significant evidence of a cutoff \cite{abr16}. It seems to provide the first evidence of a Pevatron in our Galaxy. With its current rate of particle acceleration Sgr A* cannot contribute substantially to Galactic cosmic rays but it could have been more active in the past and thereby Sgr A* is proposed as a viable alternative to supernova remnants as a source of PeV Galactic cosmic rays \cite{abr16}. However, it is not yet confirmed that whether gamma rays surrounding the galactic center is due to a single accelerator at the center or to multiple accelerators filling the region. There are also alternative interpretation of the HESS galactic centre observations (for instance see \cite{gag17}).   

The SNR origin model of cosmic rays has received some supports from the TeV gamma ray observations. If the cosmic rays are accelerated in SNRs, hadronic interactions of cosmic ray nuclei with the ambient matter/radiation will produce neutral and charged pions which in turn decay into gamma rays and neutrinos respectively. Therefore, SNRs are expected as emitter of gamma rays and neutrinos \cite{dru94, nai94, stu97, roy99}. While detection of high energy neutrinos is difficult owing to their weak interaction behavior, the theoretically predicted gamma ray fluxes in GeV to TeV energy bands from SNRs are well within the reach of modern high energy gamma ray observatories and indeed several SNRs have been detected by the modern gamma-ray observatories in TeV and GeV energies in the past fifteen years or so (\cite{aha13} and references therein). However, the evidence is only supportive but not conclusive as leptonic mechanisms such as inverse Compton scattering of thermal/ambient photons with energetic electrons also may lead to the TeV gamma-ray emission from the SNRs. In fact a leptonic scenario seems to be preferred for RXJ1713 from Fermi and HESS measurements, even if a proton sub-component cannot be excluded \cite{abd11}. 

The observation of gamma rays from SNR surrounding dense medium of molecular clouds is another handle to probe the presence of hadronic cosmic rays in the SNR \cite{aha96}. In this case gamma rays are produced due to interaction of cosmic rays those escaped from nearby SNR with molecular clouds. Gamma rays of GeV energies from some SNRs interacting with molecular clouds such as W51C \cite{abd09}, W44 \cite{abd10a, giu11}, IC 443 \cite{abd10b, tav10} and W28 \cite{aha08, abd10c, giu10} have been observed by the Large Area Telescope (LAT) on board the Fermi Gamma-ray Space Telescope and the Gamma-Ray Image Detector (GRID) on board AGILE satellite. The current generation of imaging atmospheric Cherenkov telescopes have also detected the systems in the TeV energy range \cite{aha08}. The GeV gamma-ray luminosity of the molecular clouds nearby to SNR W28 as observed from the Earth reaches $\sim 10^{36}$ erg s$^{-1}$ level \cite{abd10c}. The Inverse Compton (IC) origin of the GeV gamma rays is thus disfavored since it requires total electron energy comparable to or larger than the typical kinetic energy $\sim 10^{51}$ ergs released by a supernova explosion. Besides the strong correlation detected particularly at TeV energies between the gamma-ray intensity and dense gas is not expected for such evolved SNR \cite{abd10c}. There is an alternative hadronic explanation \cite{uch10} of observed high energy gamma rays from molecular clouds in the proximity of SNR based on the so called “crushed cloud” model \cite{bla82} in which shock-accelerated cosmic rays are trapped along with the shocked molecular cloud by the supernova blast wave and thereby form cloud shock. Subsequently re-acceleration of pre-existing cosmic rays in the molecular cloud take place. The re-accelerated cosmic rays interacting with molecular cloud produce $\pi^{o}$ mesons those decay in to high energy gamma rays.

The observed TeV gamma-ray emission in very dense molecular region also can be produced by $\sim 10$ TeV electrons through Bremsstrahlung process \cite{giu11}. But this mechanism is also ruled out as the energetic electrons are expected to be produced at early epochs of SNR and they can only marginally survive for the middle-old age SNRs because of strong synchrotron and IC cooling. So observation of TeV gamma rays from the molecular cloud instead of directly from SNR gives an added advantage and hence provide strong support to a hadronic origin of the gamma-ray emission. The characteristic spectral feature of gamma ray emission detected form W44 \cite{giu11} and IC 443 \cite{ack13} are recently explained the decay of $\pi^0$ produced in hadronic-induced interaction with the molecular clouds. Another clean signature for the hadronic acceleration in supernovae will be the observation of TeV neutrinos from SNRs. 

There are few issues in the SNR origin model of cosmic rays which are yet to be established. In particular one of the key unsettled issue is the maximum attainable energy by a cosmic ray particle in the supernova shock. If an ordinary supernova remnant is passing through a medium of density $N_{H} \; cm^{-3}$, the maximum energy that can be attained by a cosmic ray particle is \cite{lag83, fic86, bie93, ber99, ptu05}  

\begin{eqnarray}
E_{max}\simeq 200 Z \frac{\epsilon_{51}}{N_{H}}  \; TeV 
\end{eqnarray}

where the kinetic energy of ejecta is taken as $\epsilon_{51} \times 10^{51}$ ergs ($\epsilon_{51}$ is a free dimension less parameter), interstellar magnetic field is taken as $5$ $\mu$G, and the density of the ambient gas is $N_{H}$ $cm^{-3}$. The maximum energy given in the above equation is falling short the knee energy by about one order of magnitude for proton primary. Recently it has been suggested that an amplification of effective magnetic field strength at the shock may take place in young SNRs due to growth of magnetic waves induced by accelerated cosmic rays \cite{bel01}. Consequently the maximum energy achieved in SNR possibly can reach the knee energy for protons while Fe nuclei can reach upto around $10^{17}$ eV. Note that high magnetic field ($10$ to $100$ G) was observed in the inner envelopes of late-type stars \cite{bar87}. 

%Due to the low supernova rate in our galaxy, the evolution of SNRs are not easily visible from galactic observations. The detailed radio spectra of 26 compact sources in the starburst nucleus of M82 suggest mG fields at parsec scales in SNRs \cite{all98} which is consistent with the findings of $1$ Gauss typical observed magnetic fields at $10^{16}$ cm in massive star explosions (for instance Soderberg, 2005; Soderberg et al, 2005a). Even from the size-magnetic field evolution, baryon loading and energetics, using the observed radio spectra of SN 2009bb it was demonstrated (Chakraborti et al, 2011)  that a sub-population of type Ibc supernovae with mildly relativistic outflows satisfies the Hillas criterion (Hillas, 1984) to confine and accelerate cosmic rays at the highest energies.  
 
Most of the PeV energy cosmic rays are likely to be already escaped from middle/old age SNRs and therefore one cannot expect gamma rays of tens of TeV energies and above from middle/old age SNRs. Instead gamma rays from SNR surrounding molecular clouds should bear the imprints of PeV cosmic rays accelerated in the SNR, if they attained such high energies. Under the circumstances, the main objective of the present work is to probe the maximum energy attained by cosmic rays in SNR W28 (a mixed-morphology old SNR which is located at an estimated distance of $\sim 2$ kpc \cite{vel02}) through TeV gamma rays and neutrinos from the four molecular clouds interacting with the SNR W28 (one of the best examples of a cosmic-ray-illuminated cloud). In particular we would like to estimate the fluxes of TeV gamma rays from the four interacting molecular clouds considering that the maximum attainable energy of cosmic rays can be $\sim 200$TeV, the theoretical upper limit without magnetic field amplification and $Z \times 3$ PeV, the maximum energy with magnetic field amplification. By comparing our findings with the observed fluxes/flux sensitivity of planned/upcoming high energy gamma ray observatories we would examine whether present or future observations can discriminate two maximum attainable energy scenarios and thereby ultimately can resolve the maximum energy issue of SNR origin model of cosmic rays. While estimating the gamma ray flux, we consider that SNR accelerate cosmic ray protons as well as heavier nuclei with the right composition as observed from the Earth. We have also estimated the TeV neutrino flux from the molecular clouds due to decay of charged pions produced in cosmic ray induced interaction with matter of molecular cloud and explore the possibility of observing such neutrinos by IceCube experiment \cite{coe16}.

We have restricted to only the molecular clouds nearby to SNR W28, which is one of the best examples of a cosmic-ray-illuminated cloud. This is because out of the four observed high energy gamma ray emitting SNR interacting molecular cloud systems, only SNRs W28 and IC443 are both TeV and GeV gamma rays emitter whereas W51C and W44 are found to emit only in the GeV band. In the case of gamma ray emission from the molecular clouds surrounding IC 443, the GeV \cite{abd10b, har99, tav10} and TeV \cite{alb07,acc09} gamma ray emission regions are shifted from each other \cite{aha13} and thereby to model gamma ray emission from GeV to TeV range different characteristics of molecular clouds, cosmic ray propagation etc need to be adopted resulting a lot of uncertainties. 

The molecular clouds associated with W28 are studied through transitions of several tracer molecules along with CO and therefore the chemically rich environment surrounding W28 is well characterized from observations \cite{max17}. The gamma ray flux from cosmic ray illuminated molecular clouds depends on several factors including mass of the clouds. There is some uncertainties over the masses of the molecular clouds of SNR W28; at least there are two different mass profiles of the molecular clouds interacting SNR W28 \cite{aha08, nic12}. The determination of mass of molecular clouds is complex. Because $H_2$, which is supposed to primarily make up an interstellar molecular cloud, is not observable directly, surveys on molecular clouds rely trace molecules like CO which is the second most abundant molecule in molecular clouds. For the determination of cloud mass from CO as tracer of $H_2$, one first measures the velocity integrated intensity $W_{CO} \equiv \int T_{CO} dv \; K km s^{-1}$, where $T_{CO}(v)$ is the temperature of CO in the cloud and v is the velocity (local standard of rest velocity), of the $J = 1–0$ transition of $^{12}CO$ and subsequently the measurement is converted to column density of hydrogen molecules through the relation $N(H_2) = X W_{CO}$ where X denotes the so called X-factor. The X factor is calculated empirically by comparing CO data, HI data, and maps of the 100 µm dust emission \cite{dam01}. Assuming all hydrogen is molecular and cosmic rays penetrate molecular clouds freely, the X-factor is $1.5 \times 10^{20} [I(^{12}CO)/(K km/s)] (cm^{-2})$  \cite{str04}. The CO tracing method is a reasonable approach for finding masses over molecular clouds with a factor of 2 or better when averaged over a suitably large region \cite{dam01}.

The mass of of the SNR W28 associated molecular clouds estimated using the same NANTEN CO (J = 1-0) data from $W_{CO}$ integrated over the range $0 - 12$ $Km \; s^{-1}$ is found about half of that obtained from $W_{CO}$ integrated over the wider velocity range $0 - 25$ $Km \; s^{-1}$ \cite{aha08}. The mass of the W28 associated molecular clouds are also estimated using the CS (1–0) transition by the 7mm MOPRA observations \cite{nic12} and the so estimated masses are found consistent (maximum deviation less than $20 \%$) with that obtained from $W_{CO}$ integrated over the wider velocity range $0 - 25$ $Km \; s^{-1}$ \cite{nic12}. We shall consider both the mass profiles given by \cite{aha08} and \cite{nic12} and examine whether the gamma ray fluxes at high energies can be estimated with reasonable accuracy despite the uncertainty in the mass profile. In doing that we shall consider the mass composition of molecular clouds as both pure hydrogen and He molecules since it (mass composition) should lie between these two. 
 
While estimating gamma ray contribution from molecular clouds illuminated by cosmic rays from nearby SNRs, often protons are considered as accelerated particles in SNRs. However, if SNRs are the true  sites of cosmic rays, they should also emit other heavier nuclei. Cosmic rays after emission from the SNRs propagate diffusively. Since diffusion constant is not the same for protons and heavier nuclei, the mass composition at molecular clouds will differ from that at the production site. The observed cosmic ray composition at the Earth is essentially resulted after diffusive propagation. But since cosmic rays below the knee energy are confined in the galaxy, the production composition should be roughly equal to the observed composition which is not true for the case of molecular clouds illuminated by a single SNR. We, therefore, shall consider both the situation that at molecular clouds cosmic ray composition is same to what we observed at the Earth and secondly cosmic ray composition at SNR is the same to the observed one.

The plan of the paper is the following - in the next section we shall describe the methodology for evaluating the TeV gamma-ray and neutrino fluxes generated in interaction of cosmic rays accelerated in SNR with the nearby molecular clouds. In Section III we shall estimate the hadronically produced gamma-rays and neutrino fluxes from the molecular clouds interacting SNR W28 over the GeV to TeV energy range and compare our estimates with the observed gamma rays spectra. Subsequently in the same section we shall estimate gamma ray flux from molecular clouds of SNR W28 at hundreds of TeV for different maximum energy of accelerated cosmic rays and shall examine whether the future generation telescopes can observe the source (via molecular clouds) if it is a Pevatron. We shall discuss our results in Section IV and conclude finally in section V.

\section{Methodology}

The production spectrum of cosmic rays at SNR follows a power law and is given by 

\begin{equation}
 \frac{dn}{dE} =K E^{-\alpha} .
\end{equation}
where $K$ denotes the proportionality constant and $\alpha$ is the spectral index. If $\xi$ the fraction of the total energy of the supernova explosion E$_{SN}$ produces the cosmic ray particles, then the proportionality constant can be written as \cite{bha02}

\begin{eqnarray}
K = \frac{(\alpha-2)\xi E_{SN}}{E_{min}^{2-\alpha}-E_{max}^{2-\alpha}} \hspace{1cm} \mbox{if}\hspace{0.5cm} \alpha \ne 2 \\ \nonumber
  = \frac{\xi E_{SN} }{\ln(E_{max}/m_p c^2)} \hspace{0.6cm} \mbox{if}\hspace{0.5cm} \alpha = 2\;
\end{eqnarray}

where $E_{min}$ is the minimum energy and $E_{max}$ is the maximum energy attainable by a Cosmic ray particle in the SNR. After emission from the SNR, cosmic rays propagate in the interstellar medium through diffusion. The probability density of finding a cosmic ray particle at a given radius $r$ from the source is given by \cite{bha02}

\begin{equation}
 P(r) = \frac{1}{8(\pi D \tau)^{3/2} } \exp(-r^2/(4D\tau))\hspace{0.1cm}
\end{equation}

where $\tau$ is the age of the supernova explosion in seconds, $D = D_0(\frac{E/Z}{10 GeV})^{\delta}$ is the diffusion coefficient of nuclei in the Galaxy with $D_0 \sim 10^{28}$ cm$^2$s$^{-1}$ \cite{ber90, str07} and where $Z$ is the atomic number of the cosmic ray nuclei and $\delta$ is a constant having value  between $0.3$ to $0.7$ \cite{ber90}. However. the diffusion is slow in dense gaseous medium of molecular clouds \cite{orm88, aha96}. The measurement of Boron to Carbon Flux Ratio in Cosmic Rays over the rigidity range 1.9 GV to 2.6 TV by the Alpha Magnetic Spectrometer (AMS-02) on the International Space Station suggests that $\delta$ is about 0.33 \cite{agu16}. On the other hand considering  21 SNRs, those are well-studied from radio wavelengths up to gamma-ray energies, as representative for the total class of SNRs, Becker Tjus et al (2016) have recently demonstrated that the cosmic ray budget can be matched well for a diffusion coefficient that is close to $D \propto E^{0.3}$, a stronger diffusion with $\delta=0.5$ cannot reproduce the observed cosmic ray energy spectrum, particularly the high-energy (TeV) component of the spectrum, if they are originated at SNRs of the galaxy \cite{bec16}. Using equation (2), the intensity of cosmic rays at a distance $r$ from the source (assuming a point source) will be 

\begin{equation}
 \frac{dn_{CR}}{dE}(r) = KP(r)E^{-\alpha} \hspace{0.5cm} cm^{-3}GeV^{-1}.
\end{equation}

We consider the situation where a nearby dense molecular cloud is illuminated by runaway relativistic cosmic ray particles accelerated at SNR. 
The differential flux of gamma rays and neutrinos of energy $E_{\gamma/\nu}$ reaching the Earth after production in interaction of runaway relativistic cosmic ray particles with molecular clouds can be written as

\begin{eqnarray}
%\small
\frac{d\Phi_{\gamma/\nu}}{dE_{\gamma/\nu}}(E_{\gamma/\nu}) = \frac{1}{4\pi d^{2}}\frac{M_{cl}}{B m_{p}}\kappa Q_{\gamma/\nu}(E_{\gamma/\nu})
\end{eqnarray}

where $\kappa$ is a constant equal to 1 for gamma rays and equal to $1/2$ for muon neutrinos due to neutrino oscillation at large distances, $d$ is the distance between the SNR and the Earth and $M_{cl}$ is the total mass of the molecular cloud, B is the mass number of the target nuclei and $Q_{\gamma/\nu}$ is the emissivity of gamma rays/neutrinos produced in $A+B$ or $ A+p$ interaction. The $Q_{\gamma/\nu}$ has been evaluated following the prescription given by \cite{kac14, kaf14, kaf16} as outlined in Appendix. In the next section, we will estimate the fluxes of gamma rays and neutrino from the molecular clouds at the surrounding region of SNR W28 using the above expressions.

\section{Results}
The SNR W28 (G6.4−0.1) is a mixed-morphology old SNR with dimensions $50'\times 45'$, located in a region rich of dense molecular gas with average density $\ge \; 5$ cm$^{-3}$ \cite{aha08, abd10c}. W28 is now in its radiative phase of evolution and located at an estimated distance of $\sim 2$ kpc. The SNR shock radius is $\sim 12$ pc and its velocity is $\sim 80$ kms$^{-1}$ \cite{rho02}. In the framework of the dynamical model \cite{cio88} and assuming that the mass of the supernova ejecta is $\sim 1.4$M$_{\odot}$, it was concluded that the supernova explosion energy is $E_{SN} = 0.4 \times 10^{51}$ erg, age is $t_{age} = 4.4\times 10^4$ yr and initial velocity is $\sim 5500$ kms$^{-1}$ \cite{aha08, nav13}. At the surrounding region of SNR W28, four $\gamma$ ray sources in GeV and TeV energies which correlate quite well with the position of four massive molecular clouds have been observed by HESS telescopes \cite{aha08}. The clouds are HESS J1801-233, located along the north eastern boundary and HESS J1800-240A, 240B, and 240C, located to the south, outside the radio boundary \cite{aha08, abd10c}. The observed GeV-TeV gamma-ray spectrum from the molecular clouds can be explained by hadronic interactions of cosmic rays accelerated at the shocks of SNR W28 with the ambient matter of molecular clouds when a power-law spectrum of protons with a power law index 2.2 is considered \cite{han14, li10}.

According to Aharonian et al. (2008), the masses of the three clouds HESS J1801-233, HESS J1800-240A and HESS J1800-240B are $\sim 5$, $6$ and $4$ (in the unit of $10^4$M$_{\odot}$) respectively from the estimation of the NANTEN CO (J = 1-0) data. Using these values we have estimated the gamma-ray flux produced in the hadronic interaction of cosmic rays with the molecular clouds considering two maximum attainable energy scenarios, Z$\times3$ PeV, which seems achievable under an amplified magnetic field situation and 200 TeV, which is the theoretical upper limit under a normal magnetic field picture. We take the composition of molecular clouds either pure proton or pure He and the composition of cosmic rays at molecular cloud is taken the same to the observed (mixed) composition of cosmic rays at the Earth. The SNR emitted cosmic ray composition at molecular clouds is taken as the same to the observed cosmic ray composition. We have also considered the effect of background gamma ray flux produced due to interaction of diffuse cosmic ray in the galaxy with the ambient proton of molecular cloud which is estimated in the same way but replacing the cosmic ray flux from the SNR by galactic diffuse cosmic ray flux at molecular clouds.   

Our results along with the observed spectra are shown in figure 1. Here we have treated $r$ and the spectral index of accelerated cosmic rays as free parameters as the magnitudes of the parameters are not definitely known. We take a lower diffusion constant $D_0 = 5\times 10^{26}$ cm s$^{-1}$ than the standard value as suggested by \cite{aha96, aha08} with $\delta = 0.33$ as recently found by the AMS 02 \cite{agu16}. For such a choice of diffusion coefficient a single power-law energy spectrum of accelerated cosmic rays with spectral index $\alpha = -2.3$ well describes the GeV-TeV gamma ray experimental data. Average distance of the molecular clouds, HESS J1801-233, 240B and 240A are found to be 12, 30, 42 parsec respectively. The efficiency of conversion of the supernova explosion energy requires $\sim 50\%$ to match the experimental results for all the three molecular clouds. However, the efficiency will become $20\%$ if the SN explosion energy is taken the standard value of $10^{51}$ erg instead of $0.4 \times 10^{51}$ erg \cite{aha08, nav13}. If the efficiency of conversion of the supernova explosion energy is taken $10\%$, as conventionally assumed, and the diffusion coefficient and distance of molecular clouds are considered as free parameters. The estimated fluxes from all the molecular clouds consistently match the observations for $D_0 = 3.5\times 10^{26}$ cm s$^{-1}$ and with average distances of the molecular clouds, HESS J1801-233, 240B and 240A are 12, 24, 30.5 parsec respectively as shown in figure 2. For the standard diffusion coefficient D$_0 = 5\times 10^{28}$ cm s$^{-1}$ with $\delta$ unaltered at $0.33$ it is found that the estimated gamma ray fluxes (the green line in the figure 2) is far lower than the observed fluxes.

Due to their limited sensitivity in the $>10$ TeV energy domain, the current generation atmospheric Cherenkov telescopes are not in a position to test the predicted flux of gamma rays from molecular clouds at around 100 TeV where the Pevatron effect is strongly revealed. Among the most sensitive next generation gamma ray observatories, the Cherenkov Telescope Array (CTA) will cover a wide energy band $20$ GeV to $>300$ TeV \cite{ach13, ong17}. The observatory, consisting of large arrays of imaging atmospheric Cherenkov telescopes in both the southern and northern hemispheres, will provide full-sky coverage and will achieve a sensitivity improved by up to an order of magnitude compared to existing imaging Cherenkov telescopes. The full operations of the project is expected to be commenced by the middle of the next decade \cite{ong17}.   
The Large High Altitude Air Shower Observatory (LHAASO) project is another most sensitive next generation instrument, to be built at 4410 meters of altitude in the Sichuan province of China \cite{dis16, liu17}. The first phase of LHAASO will consist of a $1$ $km^{2}$ array (LHAASO-KM2A) for electromagnetic particle detectors and an overlapping $1$ $km^{2}$ array of 1146 underground water Cherenkov tanks $36$ $m^{2}$ each in size for muon detection. Besides, there will be a close-packed, surface water Cherenkov detector facility with a total area of about 78,000 $m^{2}$ (LHAASO-WCDA) and 12 wide field-of-view air Cherenkov telescopes (LHAASO-WFCTA). The LHAASO will be capable of continuously surveying the $\gamma$-ray sky for steady and transient sources from about 100 GeV to 1 PeV.  The completion of the installation is expected by the end of 2021 \cite{dis16, liu17}.
 
The 5$\sigma$ detection sensitivity of upcoming CTA detector for 1000 hrs run for an $E^{−2}$ type power-law spectrum from a point source \cite{fun13}, and the same of Large High Altitude Air Shower Observatory (LHAASO) to a Crab-like point gamma ray source for 1 year run \cite{ver16, kno16} are also shown in the figures 1 and 2 to judge whether the experiments will be able to detect the estimated gamma ray fluxes from the W28 associated molecular clouds.

\begin{figure}[t]
  \begin{center}
\includegraphics[width = 0.5\textwidth,height = 0.805\textwidth,angle=0]{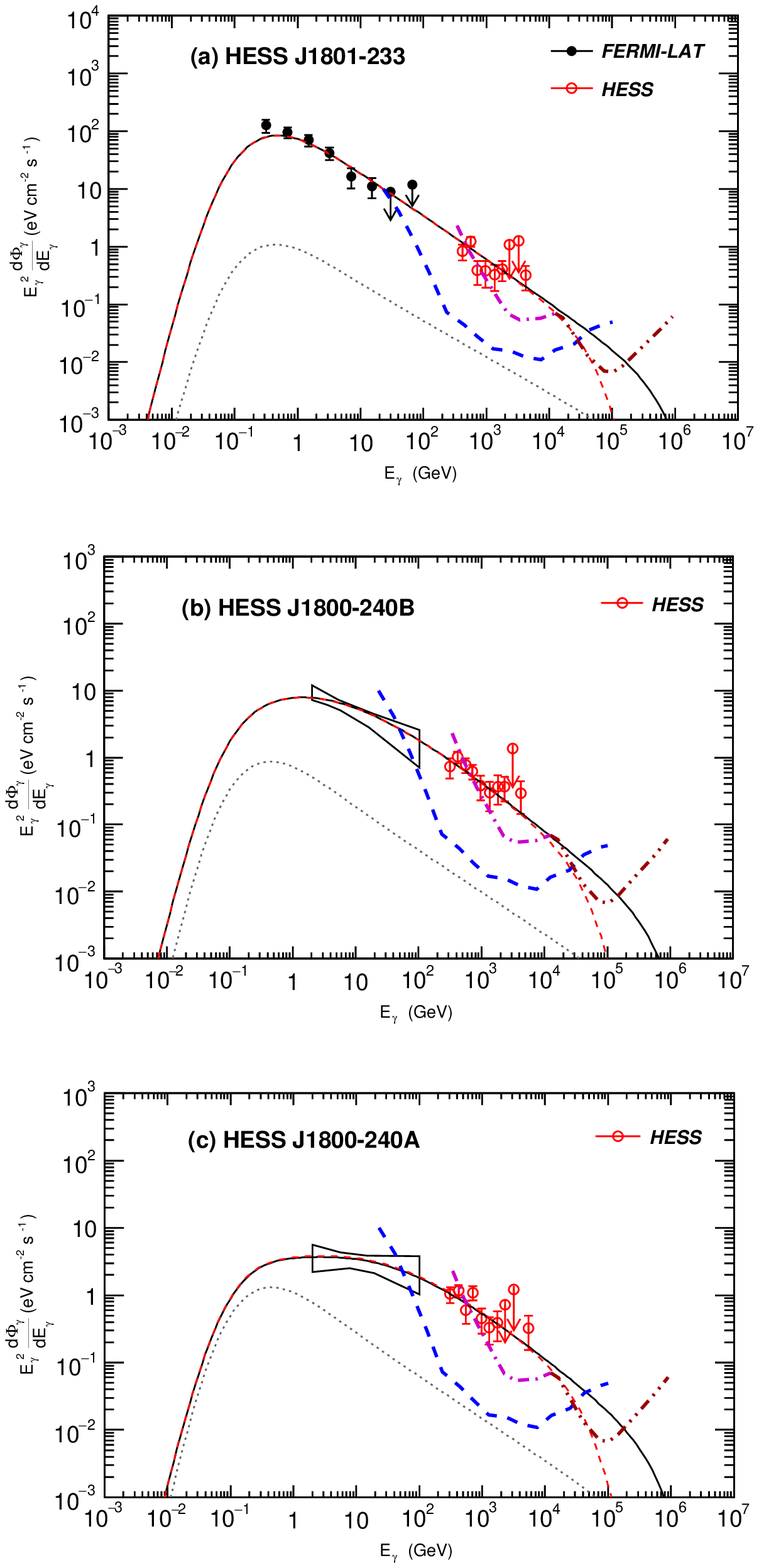}
\end{center}
%\captionsetup{margin= 30pt,font=small,labelfont=bf}
\caption{Differential energy spectrum of gamma rays reaching at the Earth from the molecular clouds assuming Aharonian et al.'s mass profile and taking diffusion coefficient equals to $D_0 = 3.5\times 10^{26}$ cm s$^{-1}$ \cite{aha08}. The $68\%$ confidence range of the LAT spectrum are shown by black lined region. The black continuous line and red dashed line indicate gamma ray fluxes when maximum attainable energy of cosmic rays is $Z \times 3$ PeV and $200$ TeV respectively. The grey dotted line denotes the background gamma ray flux from the molecular cloud due to the Cosmic rays galactic background. The blue long dashed line indicates the detection sensitivity of the CTA detector for 1000 hrs \cite{fun13}. The pink dashed single dotted line and brown dashed double dotted line indicate the detection sensitivity of the LHAASO-WCDA and LHAASO-KM2A detector respectively for 1 year \cite{ver16}.}
\label{Fig:1}
\end{figure}

\begin{figure}[t]
  \begin{center}
\includegraphics[width = 0.6\textwidth,height = 0.7\textwidth,angle=0]{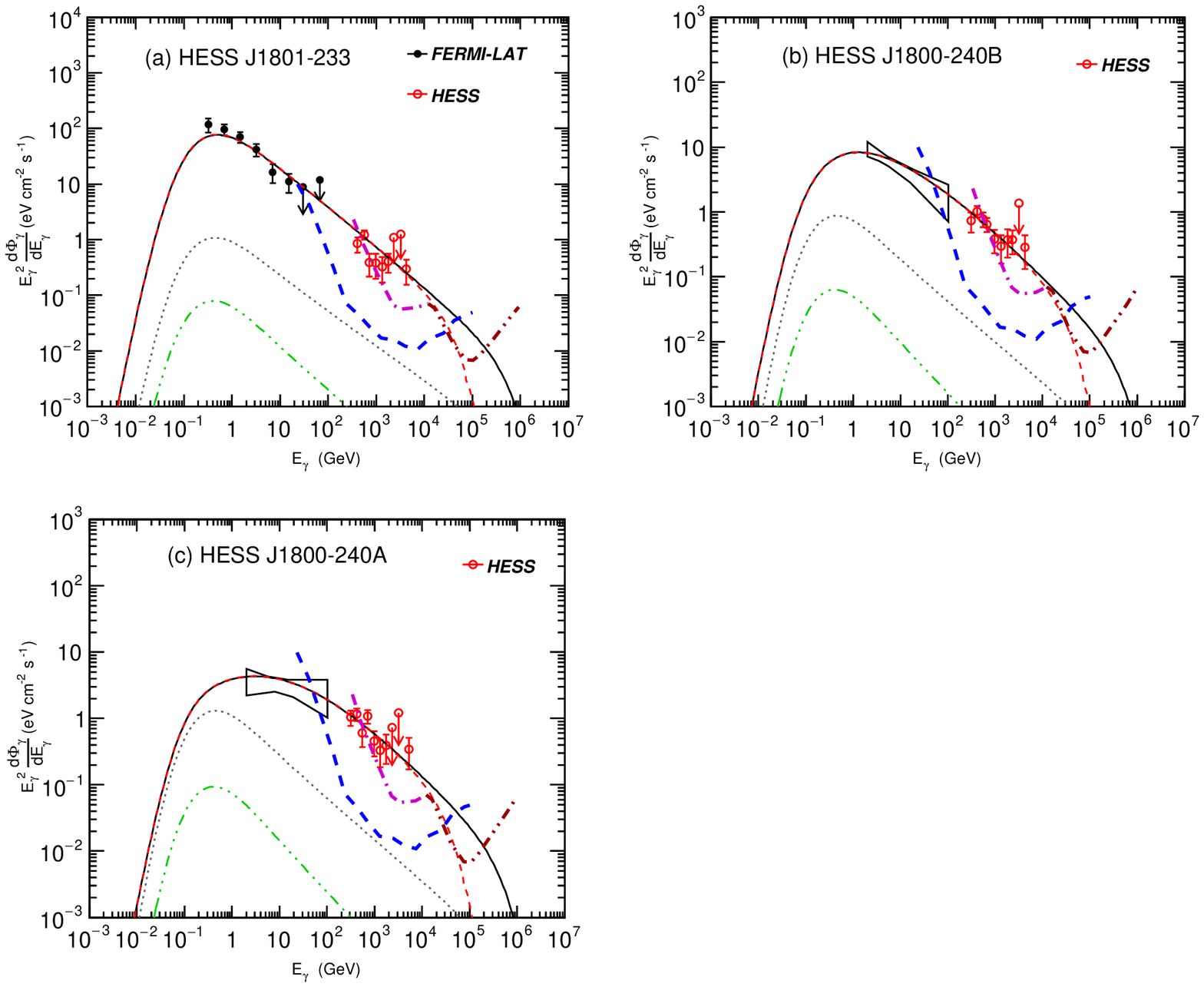}
\end{center}
%\captionsetup{margin= 30pt,font=small,labelfont=bf}
\caption{Same as Fig. 1 but for diffusion coefficient equals to $D_0 = 3.5\times 10^{26}$ cm s$^{-1}$ and $D_0 = 5\times 10^{28}$ cm s$^{-1}$ (the green dotted dashed line). }
\label{Fig:2}
\end{figure}

Nicholas et al (2012) found slightly different mass values of the four clouds HESS J1801-233, HESS J1800-240B, 240A and 240C which are given by $\sim 5$, $7.1$, $2.3$ and $1.4$ (in the unit of $10^4$ M$_{\odot}$) respectively. We repeat our analysis taking their mass profile and considering that all the four molecular clouds are correlated with the SNR W28. We found that estimated gamma ray fluxes cannot well reproduce the observed spectra for all the four clouds simultaneously and consistently. We adopt diffusion constant $D_0 = 5\times 10^{26}$ cm s$^{-1}$. The choice of the spectral index $\alpha = -2.25$,  and $\delta = 0.33$ give consistent match of the observed spectra. To match the observed spectra better we take that the molecular clouds HESS J1801-233 and HESS J1800-240B are composed of slightly heavier nuclei like He and the other two clouds HESS J1800-240A and HESS J1800-240C are mainly composed of proton. Our results along with the observed spectrum are shown in figure 3. Average distance of the molecular clouds, HESS J1801-233, 240B, 240A and 240C are found to be 12, 35, 35 and 30 parsec respectively. The efficiency of conversion of the supernova explosion energy requires $26\%$ to match the experimental results for all four molecular clouds. It is found that estimated gamma ray flux matches nearly well with the observed gamma ray flux from the three clouds, HESS J1801-233, HESS J1800-240A and HESS J1800-240C but we get slightly higher flux than the observed flux for the cloud HESS J1800-240B as shown in the figure 3. However, the mass of molecular clouds estimated using tracer molecules is uncertain by a factor of 2 or so as mentioned earlier. When we consider the mass of the cloud is $4\times10^4$M$_{\odot}$, which is within the uncertainty level, the predicted gamma ray flux from HESS J1800-240B well match the observed spectrum. If we adopt the efficiency of conversion of the supernova explosion energy as $10\%$, the matching of the estimated flux with the observation demands a relatively lower diffusion coefficient $D_0 \sim 3\times 10^{26}$ cm s$^{-1}$.

\begin{figure*}[t]
  \begin{center}
\includegraphics[width = 0.9\textwidth,height = 0.805\textwidth,angle=0]{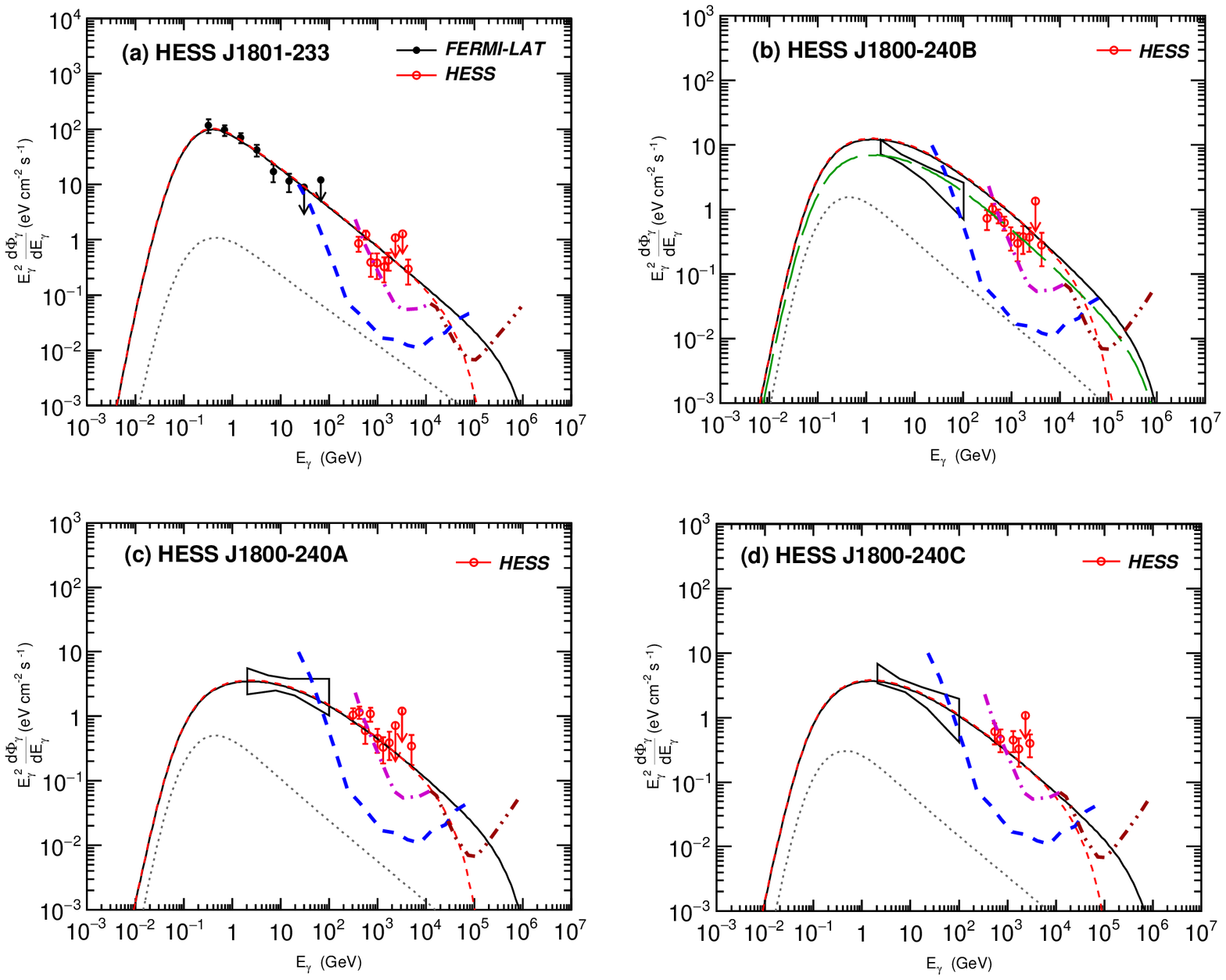}
\end{center}
%\captionsetup{margin= 30pt,font=small,labelfont=bf}
\caption{  Same as Fig. 1, but for four molecular clouds interacting SNR W28 with Nicholas's mass profiles. The diffusion coefficient is taken equals to $D_0 = 3.5\times 10^{26}$ cm s$^{-1}$ \cite{aha08}. Here the green long dashed line indicates the gamma ray flux considering mass of the cloud $4\times10^4$M$_{\odot}$,   .}
\label{Fig:3}
\end{figure*}

If cosmic ray composition at SNR (instead at molecular clouds) is taken as the same to what we observed at the Earth, then cosmic ray composition at the molecular clouds will be slightly different due to diffusion from the SNR which leads to a decrease of efficiency of conversion of supernova explosion energy into cosmic rays by $\sim 5\%$ or an increase of diffusion constant by $0.8\times 10^{26}$ cm s$^{-1}$to reproduce the observed fluxes.

We have also estimated the neutrino flux produced due to decay of charged pions in the hadronic interaction of cosmic rays accelerated in the shock of SNR W28 with the molecular clouds. The same kind of analysis using the same parameters as we did for estimating gamma ray fluxes using Aharonian's mass profile scenario has been done and the estimated neutrino fluxes from the three clouds HESS J1801-233, HESS J1800-240A and HESS J1800-240C are shown in figure 3. It is revealed from the figure 3 that the neutrino spectra from all the molecular clouds exhibit a  shoulder-like feature around 10 GeV which is absent in the gamma ray spectra. The neutrinos are produced through two different chanels unlike gamma rays, one directly through decay of charged pions and the secondly through the decay of secondary muons produced in charged pion decay. Since muons carry larger amount of energy in charged pion decay, the neutrinos produced from muon decay also carry larger energies. The neutrinos from muon decay are responsible for the shoulder-like feature in the neutrino spectra around 10 GeV. Above 10 GeV neutrino energy, the neutrino spectrum follows the primary cosmic ray spectrum similar to the case of gamma ray spectrum. Within six years of detector operation, the detection sensitivity of IceCube neutrino observatory reaches a limit of a steady flux substantially below $E^2\frac{d\phi}{dE} = 10^{-12}$ TeVcm$^{-2}$s$^{-1}$ \cite{coe16} in the northern sky for muon neutrinos ($\nu_{\mu}+\bar{\nu}_{\mu}$) having energies above 10 TeV for point-like astrophysical neutrino sources which is also shown in the figure. It is clear that the estimated fluxes are too small to be detected by IceCube. 

\begin{figure}[t]
  \begin{center}
\includegraphics[width = 0.5\textwidth,height = 0.9\textwidth,angle=0]{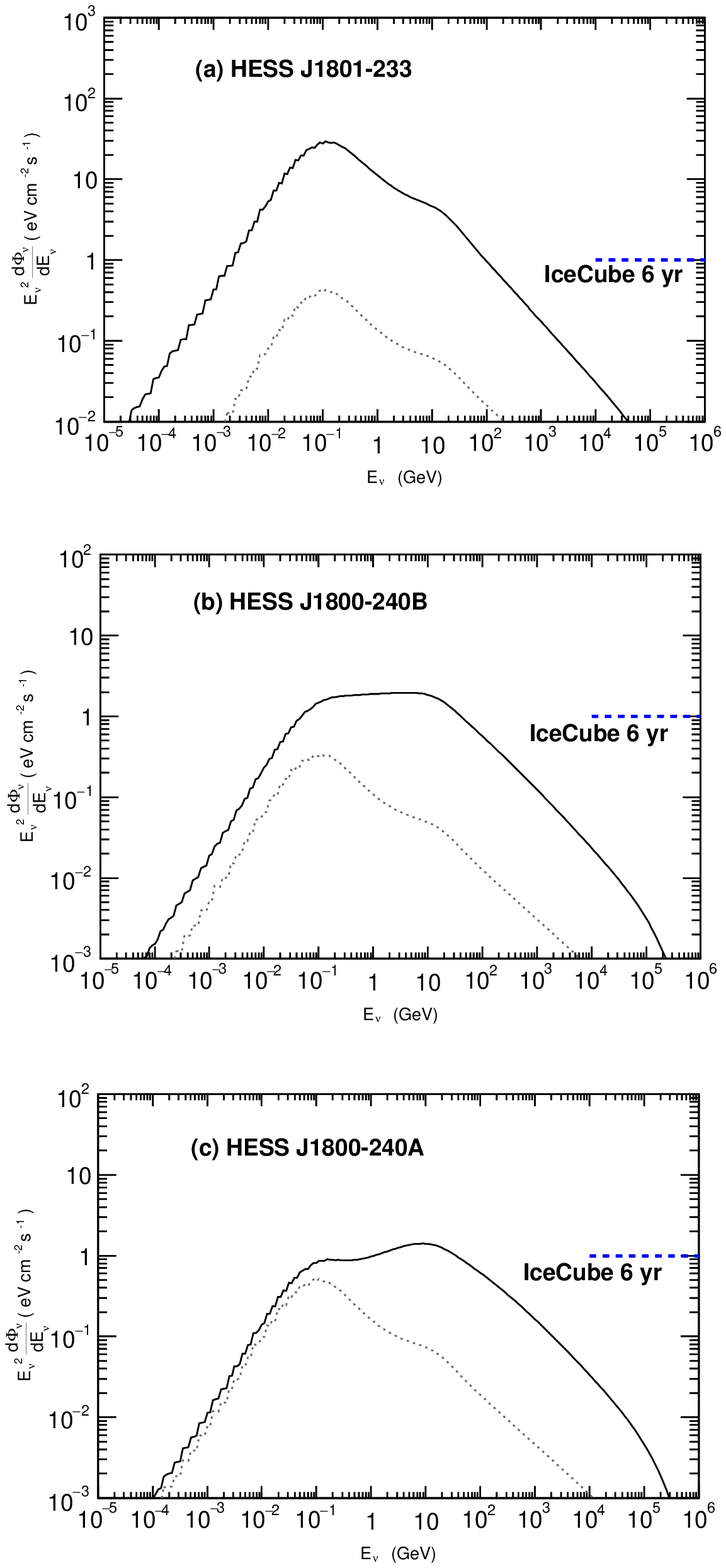}
\end{center}
%\captionsetup{margin= 30pt,font=small,labelfont=bf}
\caption{Same as the Fig.(1) but for neutrinos. The black continuous and black dotted lines indicate neutrino fluxes when maximum attainable energy of cosmic rays is $Z \times 3$ PeV and 200 TeV respectively. The blue long dashed line indicates the sensitivity of IceCube detector for 6 years of observation for point sources.}
\label{Fig:4}
\end{figure}

%\newpage
\section{Discussion}
 
The hadronic interpretation of the gamma-ray spectra of the molecular clouds illuminated by cosmic rays of SNR W28 has been advanced in several early studies \cite{abd10c, giu11, nav13, han14, li10}. In most of those earlier works protons were considered as accelerated particles in SNRs. Besides the molecular clouds are also assumed as composed of pure proton \cite{giu11, li10}. In few works the gamma-rays produced by the interaction of cosmic ray protons with ambient hydrogen is scaled by a factor of $1.84$ to account for helium and heavy nuclei in the CR composition and target material \cite{abd10b, han14} under the assumption that the composition of molecular clouds is the same as the composition of interstellar medium of our galaxy. In the present work we have assumed that the SNR W28 emits cosmic rays with proper (observed) abundances and we have considered the composition of molecular clouds as both pure proton and pure He. So whatever may be the true composition of molecular clouds, the spectral shape will remain the same, only the conversion efficiency will lie between what we obtained for proton and He .      

If the explosion energy is taken $E_{SN} = 0.4\times 10^{51}$ ergs \cite{nav13}, for the mixed primary cosmic ray composition the efficiencies of conversion of the supernova explosion energy respectively require $50\%$ to reproduce observed fluxes. The required efficiency of conversion of the supernova explosion energy into cosmic rays is quite high and difficult to achieve in an ordinary SNR. Instead if typical explosion energy of the SNR of $\sim10^{51}$ erg is considered then the conversion efficiencies will become $20\%$ which seems reasonable though still at higher side. 

Another relevant issue is the diffusion coefficient. The slope $\delta$ of the diffusion coefficient is well constrained by observations: the fitting of the recent AMS-02 measurement of Boron to Carbon Flux Ratio in Cosmic Rays above 65 GV gives $\delta \sim 0.33$ \cite{agu16}. On the other hand the cosmic rays originated in SNRs can fulfill the cosmic ray budget in the galaxy well if $\delta \sim 0.3$ \cite{bec16}. The present analysis does not impose any restriction on $\delta$ because both $\delta$ and the spectral power index $\alpha$ of SNR accelerated cosmic rays are unknown parameters. The gamma ray observation only restricts their total sum $\alpha + \delta$. So any change in $\delta$ can be compensated by corresponding changes in $\alpha$ and vice versa. We have, therefore, taken $\delta = 0.33$ as obtained by AMS-02 and found that $\alpha \sim 2.3$ reproduces the gamma ray spectral slope. Assuming the widely accepted $10\%$ conversion efficiency it is found that $D_0$ has to be $ \sim 3\times 10^{26}$ cm s$^{-1}$ for consistent match of the estimated fluxes with the observed fluxes for all the molecular clouds. The derived $D_0$ is about two order smaller than the conventional value of $D_0$ for interstellar propagation \cite{ber90, str07} as also inferred from the recent AMS-02 measurement \cite{yua17}. However, such a small $D_0$ is not unexpected in dense medium \cite{orm88, aha96} which is probably due to higher magnetic field (and hence smaller Larmor radius) in molecular clouds.               

Our main objective was to investigate the effect of maximum attainable energy of cosmic rays on gamma ray flux from the molecular clouds. We obtained gamma ray fluxes from the molecular clouds for two scenarios of the maximum energy of cosmic rays in SNRs: $Z\times3\times10^{15}$ eV, which seems achievable under an amplified magnetic field situation and $2\times10^{14}$ eV, which is the theoretical upper limit under a normal magnetic field picture. If we denote the ratio of estimated gamma ray flux from molecular clouds correspond to the maximum energy $3$ PeV to that correspond to the maximum energy of 200 TeV by $\eta$, we noticed that $\eta$ is about 2 around 30 TeV that increases to about 12.8around 100 TeV. Below 10 TeV, however, the ratio is nearly one. This finding is robust, irrespective of the different choice of mass profiles of the molecular clouds and other free parameters. 

The most relevant question is whether the stated two scenarios of maximum attainable energy of cosmic rays in SNRs can be discriminated by observations. The ongoing ground-based gamma-ray telescopes do not have the required sensitivity to discriminate the stated two scenarios as may be noticed from the figures 1 and 2. The upcoming CTA experiment will have the sensitivity for 5$\sigma$ detection of the flux levels from the molecular clouds of SNR W28 around 30-40 TeV in 1000 hours observations, if cosmic rays in W28 attained energy up to the knee energy. The planned LHAASO experiment has the best probability to observe the pevatron on gamma ray flux from the molecular clouds of W28. The large detector array (KM2A) of LHAASO should be able to detect the gamma ray flux around 100 TeV from the molecular clouds provided the cosmic rays were accelerated up to 3 PeV energy with high significance from just 1 year of observation. 

\section{Conclusion}

The maximum energy achievable at SNR shocks and acceleration efficiency are the two main key questions of the proposed ASTRO-H space observatory \cite{aha14}. Theoretically cosmic rays can be accelerated to PeV energies in SNRs provided the effective magnetic field in SNR environment is amplified by about two orders due to growth of magnetic waves induced by accelerated cosmic rays. It is worthwhile to mention that the observations of SNRs in other galaxies suggest for magnetic field of mG and higher in the parsec distance scales and hence theoretically SNRs can accelerate cosmic rays to PeV energies and beyond. But so far no observational proof has been found that SNRs can accelerate cosmic rays up to the “knee” energy. Note that among the SNRs detected so far in very high energy gamma-rays which include RXJ1713, HESS J1641-463, Cas A, W28, IC433, none of them are currently accelerating cosmic rays to PeV energies, while it may have been the case in the past. The molecular clouds located nearby of SNRs provide opportunity to trace the run-away PeV cosmic rays via the secondary gamma rays produced in interaction of cosmic rays with molecular clouds. We here examined the implications of two different maximum energies that may be achieved by cosmic rays in SNR W28 on TeV gamma rays and neutrinos from the four molecular clouds,  HESS J1800-240A, J1800-240B, J1800-240C and HESS J1801-233 illuminated by W28 emitted cosmic rays.

Our findings suggest that the gamma ray flux above about 30 TeV will be significantly higher if cosmic rays attain PeV energies in comparison to that corresponds to the cosmic rays of maximum energy 200 TeV.  The gamma ray flux level at such high energies is detectable at $5\sigma$ level by the upcoming CTA experiment with about 1000 hours exposure and also by the planned LHAASO (KM2A) telescope with about 1 year exposure. However, even if CTA observe gamma rays around 30 TeV from molecular clouds adjacent to SNR W28 one cannot definitely conclude that the SNR is a Pevatron because 30 TeV gamma rays are produced from cosmic ray protons with energies of around 300 TeV. The $5\sigma$ detection of the molecular clouds of SNR W28 around 100 TeV in future by the LHAASO (KM2A) telescope would lead to the strong conclusion that the SNR W28 is a Pevatron and thereby resolve the observational issue of maximum attainable energy of cosmic rays in SNR.  

%\section{Appendix}

\section{Acknowledgments}
The authors would like to thank an anonymous reviewer for insightful comments and very useful suggestions that helped us to improve and correct the manuscript.  The work of P. B. is supported by the UGC (India) under the Award No. F.17-88/98 (SA-1).
%\end{document}

\appendix
\section{Emissivity of gamma rays/neutrinos}
If the molecular cloud is composed of pure protons then the emissivity of $\pi^0$ mesons per target atom produced in interaction of cosmic ray projectile of mass number $A$ with the target proton is given by \cite{anc07, ban17} 

\begin{eqnarray}
%\small
Q_{\pi^{0}}^{Ap}(E_{\pi^{0}}) = c \int_{E_{N}^{th}(E_{\pi^{0}})}^{E_{N}^{max}}\frac{dn_{A}}{dE_{N}}\frac{d\sigma_{A}}{dE_{\pi^{0}}}(E_{\pi^{0}},E_{N})dE_{N}
\end{eqnarray}

where $E_{N}$ is the energy per nucleon, $d\sigma_{A}/dE_{\pi}$ is the differential inclusive cross section for the production of a pion with energy $E_{\pi}$ in the lab frame by the stated process and $E_{N}^{th}(E_{\pi})$, the threshold energy per nucleon is determined through kinematic considerations required to produce a pion with energy $E_{\pi}$. The parametrization of the differential cross section for the inclusive cross section of $A+p$ interaction used here is given below \cite{anc07, sup16} 

\begin{equation}
%\small
\frac{d\sigma_{A}}{dE_{\pi}}(E_{\pi},E_{N}) \simeq \frac{A\sigma_{0}}{E_{N}}F_{\pi}(x,E_{N})
\end{equation}
where $x = E_{\pi}/E_{N}$. The inelastic cross section of p-p interactions ($\sigma_{0}$) is given by \cite{kel06} 
\begin{eqnarray}
%\small
\sigma_{0}(E_{N}) = 34.3+1.88L+0.25L^2 \, mb
\end{eqnarray}
where $L = \ln(E_{N}/TeV)$.

We use the empirical function for the energy distribution of secondary pions that well describes the simulation results obtained with the SIBYLL code \cite{kel06}. 

Due to decay of $\pi^0$ mesons, the resulting gamma ray emissivity is given by 

\begin{eqnarray}
%\small
Q_{\gamma}^{Ap}(E_{\gamma}) = 2\int_{E_{\pi}^{min}(E_{\gamma})}^{E_{\pi}^{max}}\frac{Q_{\pi^{0}}^{Ap}(E_{\pi})}{(E_{\pi}^2-m_{\pi}^2)^{1/2}}dE_{\pi}
\end{eqnarray}
where $m_{\pi}$ is the mass of a pion and  $E_{\pi}^{min}(E_{\gamma}) = E_{\gamma} + m_{\pi}^2/(4E_{\gamma})$, is the minimum energy of a neutral pion required to produce a gamma ray photon of energy $E_{\gamma}$ .

If the composition of molecular cloud is heavier than proton having mass number $B$, the emissivity $Q_{\pi}^{AB}(E_{\pi})$ of $\pi^0$ mesons per target atom  produced due to hadronic interaction of cosmic rays coming from nearby SNR with target cloud nuclei can be obtained from equation (6) by replacing $\frac{d\sigma_{A}}{dE_{\pi}}$ with nucleus-nucleus inclusive pion production cross section as given below \cite{kac14, kaf14, kaf16} 

\begin{equation}
%\small
\frac{d\sigma_{\pi}^{AB}}{dE_{\pi}}(E_{\pi},E_{N}) \simeq \frac{w_{AB}}{2} \times \frac{\sigma_{inel}^{AB}}{E_{N}}F_{\pi}(x,E_{N})
\end{equation}

where $F_{\pi}(x,E_{N})$ is same as equation (9), $w_{AB}$ is the number of wounded nucleons and $\sigma_{inel}^{AB}$ is inelastic cross section in nucleus-nucleus interaction. In addition the threshold energy per nucleon has to be replaced by $E_{N}^{th}(E_{\pi}) = E_{\pi}+(\frac{1}{A}+\frac{1}{B})m_{\pi}+\frac{m_{\pi}^2}{2m_pAB} $ \cite{kaf16}.

For nucleus-nucleus interaction, the reaction cross section reads \cite{kaf14, sih93} 
\begin{eqnarray}
%\small
\sigma_{R} = \sigma_{R_{0}}[A^{1/3}+B^{1/3}-\beta_{0}(A^{-1/3}+B^{-1/3})]^2
\end{eqnarray}

where, $A$ and $B$ are the projectile and the target mass numbers respectively and $\sigma_{R_{0}} = \pi r_0^2 \approx 58.1$ mb with $r_0 = 1.36$ fm. If the projectile is proton, then the coefficient $\beta_0 = 2.247 - 0.915(1+B^{-1/3})$ and for projectile different from proton $\beta_0 = 1.581 - 0.876(A^{-1/3}+B^{-1/3})$. The energy dependence of the cross section at very high energies can be described by modifying the above formula as \cite{kaf14} 

\begin{eqnarray}
%\small
\sigma_{inel}^{AB}(E_N) = \sigma_{R}(A,B)*\zeta (E_{N})
\end{eqnarray}

where, $E_N$ is the energy per nucleon of the projectile and the function $\zeta (E_N)$ is given by \cite{kaf14} 

\begin{eqnarray}
%\small
\zeta (E_N) = 1+ \log \left( max \left[1,\frac{\sigma_{0}(E_N)}{\sigma_{0}(E_N^0)} \right] \right)
\end{eqnarray}

where $\sigma_{0}(E_N)$ is the $pp$ inelastic cross section and $E_N^0 = 10^3$ GeV. 

The number of wounded nucleons in nucleus-nucleus interaction can be written as (following the ``wounded nucleons" model) \cite{kaf14} 

\begin{eqnarray}
%\small
w_{AB} = \frac{A\sigma_{pB}+B\sigma_{pA}}{\sigma_{AB}}
\end{eqnarray}

where A and B are two nuclei with mass numbers $A$ and $B$, $\sigma_{AB}$ is the inelastic cross section of the reaction $A+B$, $\sigma_{pA}$ and $\sigma_{pB}$ are the nucleon(proton)-nucleus $A$ or $B$ inelastic cross sections. The emissivity of gamma rays $Q_{\gamma}^{AB}(E_{\gamma})$ due to decay of $\pi^0$ mesons produced in nucleus-nucleus interaction can be obtained from equation (10) by replacing $Q_{\pi^{0}}^{Ap}(E_{\pi})$ with $Q_{\pi^{0}}^{AB}(E_{\pi})$.

When cosmic ray nuclei interact with the nuclei of the molecular clouds, charged pions $\pi^{\pm}$ are also created. The charged pions decay into neutrinos and the emissivity of such neutrinos per target nuclei produced in the above stated  process can be written as 
\begin{eqnarray}
%\small
Q_{\nu}^{Ap}(E_{\nu}) = c \int_{E_{\nu}}\frac{dn_{A}}{dE_{N}}\frac{d\sigma_{\nu}^{Ap}}{dE_{\nu}}(E_{\nu},E_{N})dE_{N}
\end{eqnarray}

where the inclusive cross section for neutrino production is

\begin{equation}
%\small
\frac{d\sigma_{\nu}^{Ap}}{dE_{\nu}}(E_{\nu},E_{N}) \simeq \frac{A\sigma_0}{E_{N}}F_{\nu}(x,E_{N})
\end{equation}

and $F_{\nu}(x,E_{N})$ is the total neutrino production spectrum in all flavor and $x= E_{\nu}/E_N$.

In direct decay of charged pions, produced secondary muons subsequently decay $\mu \rightarrow e\nu_\mu \nu_e $ into electrons/positrons and neutrinos. The spectra of electrons is well described by the following function \cite{kel06} 

\begin{eqnarray}
%\small
F_{e}(x,E_{N}) = B_e \frac{(1+k_e(\ln x)^2)^3}{x(1+0.3/x^{\beta_e})}(-\ln x)^5
\end{eqnarray}

where

\begin{eqnarray}
%\small
B_e = \frac{1}{69.5+2.65L+0.3L^2},
\end{eqnarray}

\begin{eqnarray}
%\small
\beta_e = \frac{1}{(0.201+0.062L+0.00042L^2)^{1/4}},
\end{eqnarray}

\begin{eqnarray}
%\small
k_e = \frac{0.279+0.141L+0.0172L^2}{0.3+(2.3+L)^2}.
\end{eqnarray}

where $L=\ln(E_{N}/TeV)$ and $x= E_e/E_N$. The same function can be used to describe $F_{\nu_{\mu}^{(2)}}(x,E_{N})$, the spectrum of muonic neutrino from the decay of muon, by replacing $x= E_{\nu_{\mu}}/E_N$. Thereby in the direct decay of pions $\pi \rightarrow \mu \nu_\mu$, the spectrum of muonic neutrino can be described as \cite{kel06} 

\begin{eqnarray}
%\small
F_{\nu_{\mu}^{(1)}}(x,E_{N}) = B' \frac{\ln(y)}{y}\left(\frac{1-y^{\beta'}}{1+k'y^{\beta'}(1-y^{\beta'})}\right)^4 \\ \nonumber
\left[\frac{1}{\ln(y)}-\frac{4\beta'y^{\beta'}}{1-y^{\beta'}}- \frac{4k'\beta'y^{\beta'}(1-2y^{\beta'})}{1+k'y^{\beta'}(1-y^{\beta'})}\right]
\end{eqnarray}
where $x= E_{\nu_{\mu}}/E_N$, $y= x/0.427$, 

\begin{eqnarray}
%\small
\beta' = \frac{1}{1.67+0.111L+0.0038L^2},
\end{eqnarray}

\begin{eqnarray}
%\small
B' = 1.75+0.204L+0.010L^2,
\end{eqnarray}

\begin{eqnarray}
%\small
k' = 1.07-0.086L+0.002L^2
\end{eqnarray}
At $x = 0.427$, the spectrum of $F_{\nu_{\mu}^{(1)}}$ exhibits sharp cut off. So the total spectrum of muon neutrinos is $F_{\nu} = F_{\nu_{\mu}^{(1)}}+F_{\nu_{\mu}^{(2)}}$.

\end{document}